\documentclass{article}
\usepackage{latexsym}
\usepackage{graphicx}
\begin{document}
\begin{center}
\textbf{\LARGE On Scalar and Vector Potentials for the Nonlinear \\ 
\vspace{.3cm} Electromagnetic Forces}
\end{center}

%\begin{large}

\begin{center}
W. Engelhardt\footnote{ Electronic address: 
wolfgangw.engelhardt@t-online.de}
\end{center}

\begin{center}
retired from: Max-Planck-Institut f\"{u}r Plasmaphysik, Garching, Germany
\end{center}

\vspace{.5cm}

\textbf{\large Abstract}
 
The potential concept that is successful in classical electrodynamics should 
also be applicable to the nonlinear electromagnetic forces acting on matter. 
The obvious method of determining these potentials should be provided by 
Helmholtz's theorem. It is found, however, that the theorem fails in most 
practical instances. Other methods to find the potentials -- as pursued in plasma physics -- are 
examined and found to yield functions which 
depend on the chosen coordinate system. Thus they cannot be considered as 
invariant potentials from which physical forces may be derived. Practical 
consequences of these mathematical findings are discussed.

\vspace{.5cm}

\textbf{\large Keywords}

\begin{enumerate}
\item Lorentz force
\item Maxwell's equations
\item Helmholtz's theorem
\item magnetized plasmas
\item ideal and resistive magnetohydrodynamics
\item magnetic plasma confinement
\end{enumerate}
%\vspace{.5cm}

\textbf{\large P.A.C.S.:}  41.20.Cv, 41.20.Gz, 45.20.da, 47.65.-d, 52.25.Xz, 52.30.Cv, 52.55.-s, 52.65.Kj
\vspace{.5cm}

\textbf{\large 1 Introduction }

In classical electrodynamics the force density acting on a volume element, 
which carries charge and current, is derived from the divergence of 
Maxwell's stress tensor. We have the well known result \cite{1} which may also be 
derived directly from the Lorentz force:
\begin{equation}
\label{eq1}
\vec {F}=\rho \;\vec {E}+\vec {j}\times \vec {B}
\end{equation}
Both the electric and the magnetic force are nonlinear, since the 
electromagnetic field itself depends on the sources $\rho $ and $\vec {j}$ 
according to Maxwell's linear equations (in vacuo):
\begin{equation}
\label{eq2}
div\,\vec {E}=\rho \mathord{\left/ {\vphantom {\rho {\varepsilon _0 }}}\right.\kern-\nulldelimiterspace}{\varepsilon _0 }
\end{equation} 
\begin{equation}
\label{eq3}
rot\,\vec 
{E}=-\frac{\partial \vec {B}}{\partial t}
\end{equation} 
\begin{equation}
\label{eq4}
rot\,\vec {B}=\mu _0 {\kern 
1pt}\vec {j}+\frac{1}{c^2}\frac{\partial \vec {E}}{\partial t} 
\end{equation} 
\begin{equation}
\label{eq5}
div\,\vec {B}=0
\end{equation} 
It is convenient to introduce a scalar and a vector potential 
\begin{equation}
\label{eq6}
\vec {E}=-\nabla \phi -\frac{\partial \vec {A}}{\partial t}\;,\quad \quad 
\vec {B}=rot\,\vec {A}
\end{equation}
from which the fields may be derived. The ansatz (\ref{eq6}) satisfies equations (\ref{eq3}) 
and (\ref{eq5}) automatically. Upon substitution into (\ref{eq2}) and (\ref{eq4}) it leads to two 
second order equations which may be solved by known mathematical methods.

In view of the success of the potential method it is somewhat surprising 
that textbooks do not make use of it by applying it directly to (\ref{eq1}). This 
should be possible, since, according to Helmholtz's theorem \cite{2}, any vector 
field may be decomposed to the form:
\begin{equation}
\label{eq7}
\vec {F}=\nabla p+rot\,\vec {u}\;,\quad div\,\vec {u}=0
\end{equation}
This way the force $\vec {F}$ is expressed in terms of the scalar potential 
$p$ and the vector potential $\vec {u}$. The advantage of the decomposition 
becomes obvious when we want to balance the electromagnetic forces with 
mechanical forces inside conductors. In plasma confinement physics, e.g., 
one would like to balance the mechanical pressure gradient with the magnetic 
force:
\begin{equation}
\label{eq8}
\vec {j}\times \vec {B}=\nabla p
\end{equation}
This requires finding a magnetic field configuration such that the 
rotational part of the cross product vanishes. The electric force in (\ref{eq1}) is 
of less practical importance, since the charge distribution is usually not 
even known. Instead one calculates the electric field from Ohm's law and 
determines subsequently the charge density from (2). In this case, however, 
Ohm's law for a moving conductor in the form
\begin{equation}
\label{eq9}
\vec {E}+\vec {v}\times \vec {B}=\eta \,\vec {j}
\end{equation}
requires us also to balance the gradient part of the cross product with the 
gradient part of the electric field in steady state. 

From these examples it should be clear that it is worthwhile to explore the 
possibility of expressing a cross product as the sum of a gradient and a 
curl. In Section 2 we take a closer look at Helmholtz's theorem and find 
certain limitations of its validity. It appears that it cannot always be 
used to solve first order partial differential equations of the type (\ref{eq8}) or 
(\ref{eq9}) in a similar way as the ansatz (\ref{eq6}) satisfied the first order equations 
(\ref{eq3}) and (\ref{eq5}). In Section 3 we examine the usual methods of solving first 
order equations of type (\ref{eq8}) or (\ref{eq9}). The surprising result is that the scalar 
potentials turn out to depend on the chosen coordinate system and can, 
therefore, not represent physical quantities such as pressure or 
electrostatic potential. In Section 4 we discuss some physical consequences 
of our findings.

\vspace{.5cm}

\textbf{\large 2 Restrictions on the validity of Helmholtz's theorem}

Helmholtz's theorem is proven \cite{2} in the infinite domain by referring to the 
uniqueness theorem which applies to the solution of Poisson's equation. When 
we take the divergence of the defining equation (\ref{eq7})
\begin{equation}
\label{eq10}
\Delta p=div\,\vec {F}
\end{equation}
we obtain a Poisson equation which has the unique solution
\begin{equation}
\label{eq11}
p\left( {\vec {x}} \right)=-\frac{1}{4\pi }\int\!\!\!\int\!\!\!\int 
{\frac{div'\,\vec {F}\left( {\vec {x}'} \right)}{\left| {\vec {x}-\vec {x}'} 
\right|}} \,d^3x'
\end{equation}
subject to the boundary condition $p\left( \infty \right)=0$. Similarly, 
taking the rotation of (\ref{eq7}) one obtains the vector Poisson equation 
\begin{equation}
\label{eq12}
\Delta \vec {u}=-rot\,\vec {F}
\end{equation}
with the solution
\begin{equation}
\label{eq13}
\vec {u}\left( {\vec {x}} \right)=\frac{1}{4\pi }\int\!\!\!\int\!\!\!\int 
{\frac{rot'\,\vec {F}\left( {\vec {x}'} \right)}{\left| {\vec {x}-\vec {x}'} 
\right|}} \,d^3x'
\end{equation}
subject to the boundary condition $\vec {u}\left( \infty \right)=\vec {0}$. 
Substitution of the potentials (\ref{eq11}) and (\ref{eq13}) into (\ref{eq7}) should yield the force 
field $\vec {F}\left( {\vec {x}} \right)$ again. 

At first sight this conjecture is not expected to be true for a simple 
reason: The solutions (\ref{eq11}) and (\ref{eq13}) are completely analogous to the 
solutions for the electromagnetic potentials which arise from Poisson 
equations when the ansatz (\ref{eq6}) is substituted into (\ref{eq2}) and (\ref{eq4}) in the static 
case. Whereas the sources $\rho $ and $\vec {j}$ are localized, the 
potentials and the fields derived from them are finite also in the region 
outside the sources. The force (\ref{eq1}), however, is proportional to the sources 
and, hence, vanishing outside the conductors. It would be surprising, if 
integrals of the type (\ref{eq11}) and (\ref{eq13}) would yield finite electromagnetic 
fields outside the source region, but vanishing electromagnetic forces in 
agreement with (\ref{eq1}). In fact, this is not the case in general, but holds 
only, when the force vanishes on the surface of the conductor where charge 
and current density are still finite. 

\begin{figure}[htbp]
\centerline{\includegraphics[width=5.0in,height=2.5in]{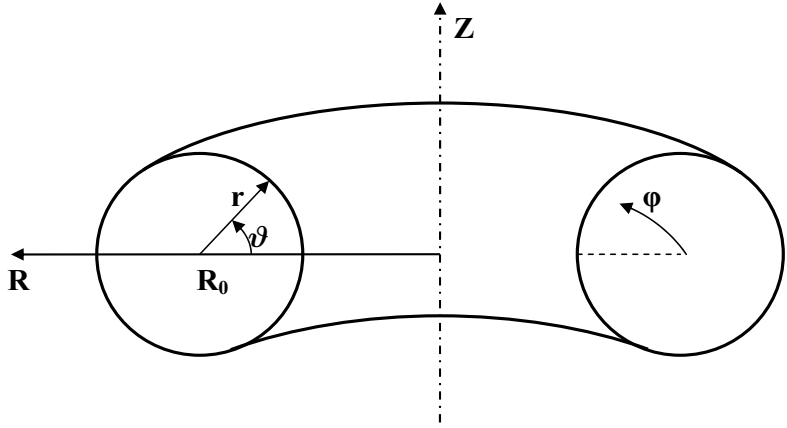}}
\caption{ Axisymmetric toroidal solenoid}
\label{torus}
\end{figure}

\begin{figure}[htbp]
\centerline{\includegraphics[width=5.0in,height=4.0in]{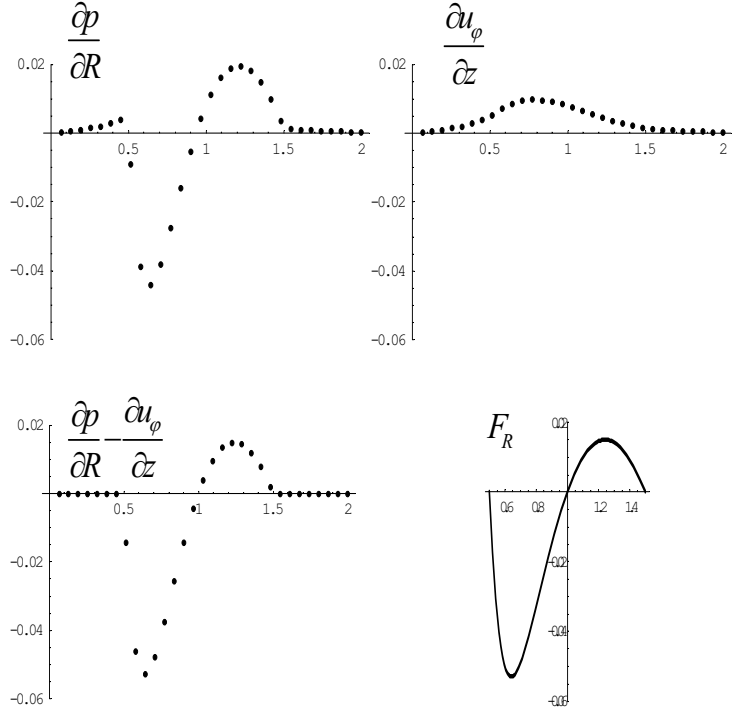}}
\caption{Radial magnetic force $F_R =j_\vartheta \,B_\varphi $ represented 
by potentials}
\label{pot1}
\end{figure}

In order to demonstrate this we consider the magnetic force in its 
decomposition
\begin{equation}
\label{eq14}
\vec {j}\times \vec {B}=\nabla p+rot\,\vec {u}
\end{equation}
and choose a toroidal geometry as sketched in Figure \ref{torus}. 
The force in the direction of the minor radius has two contributions:
\begin{equation}
\label{eq15}
F_r =j_\vartheta \,B_\varphi -j_\varphi \,B_\vartheta 
\end{equation}
Firstly, we consider a magnetic field which points only in toroidal 
direction and vanishes at the boundary $r=r_b $:
\begin{equation}
\label{eq16}
B_\varphi =C\frac{r_b^2 -r^2}{2\,R}\;,\quad R=R_0 -r\cos \vartheta 
\end{equation}
Insertion into the static equation (4) gives the current density components:
\begin{equation}
\label{eq17}
j_\vartheta =\frac{C}{\mu _0 }\frac{r}{R}\;,\quad j_r =0
\end{equation}
so that the magnetic force has the components:
\begin{equation}
\label{eq18}
F_r =\frac{C^2}{\mu _0 }\frac{r\left( {r_b^2 -r^2} \right)}{2\,R^2}\;,\quad 
F_\vartheta =0
\end{equation}
Now we must take the divergence of this force to obtain the Laplacian of the 
scalar potential:
\begin{equation}
\label{eq19}
\Delta p=\frac{C^2}{\mu _0 \,R\,r}\frac{\partial }{\partial r}\left( 
{\frac{R\,r^2\left( {r_b^2 -r^2} \right)}{2\,R^2}} \right)
\end{equation}
Substitution into (\ref{eq11}) yields the potential:
\begin{eqnarray}
\label{eq20}
 p\left( {R,\,Z} \right)=\frac{-C^2}{4\pi \,\mu _0 }\int\limits_0^{2\pi } 
{\int\limits_0^{2\pi } {\int\limits_0^{r_b } {\frac{1}{R'\,r}\frac{\partial 
}{\partial r}\left( {\frac{R'\,r^2\left( {r_b^2 -r^2} \right)}{2\,R'^2}} 
\right)} } } \frac{R'\,r\,dr\,d\vartheta \,d\varphi }{\left| {\vec {x}-\vec 
{x}'} \right|} \nonumber \\ 
  \left| {\vec {x}-\vec {x}'} \right|=\sqrt {R^2+R'^2-2\,R\,R'\cos \varphi 
+\left( {Z-r\sin \vartheta } \right)^2} \;,\quad R'=R_0 -r\cos \vartheta 
\end{eqnarray}
Taking the rotation of (\ref{eq18}) one obtains with (\ref{eq13}) for the toroidal component 
of the vector potential:
\begin{equation}
\label{eq21}
u_\varphi \left( {R,\,Z} \right)=\frac{-C^2}{4\pi \,\mu _0 
}\int\limits_0^{2\pi } {\int\limits_0^{2\pi } {\int\limits_0^{r_b } 
{\frac{1}{r}\frac{\partial }{\partial \vartheta }\left( {\frac{\,r\left( 
{r_b^2 -r^2} \right)}{2\,R'^2}} \right)} } } \frac{\cos \varphi 
\,\,R'\,r\,dr\,d\vartheta \,d\varphi }{\left| {\vec {x}-\vec {x}'} \right|}
\end{equation}
Rather than trying to evaluate these integrals analytically, we have 
calculated them numerically and compared the result for the force component 
in the midplane $Z=0$
\begin{equation}
\label{eq22}
F_R =\frac{\partial p}{\partial R}-\frac{\partial u_\varphi }{\partial Z}
\end{equation}
with the analytic expression resulting from (\ref{eq18}):
\begin{equation}
\label{eq23}
F_R =\frac{C^2}{\mu _0 }\frac{\left( {R-R_0 } \right)\left( {r_b^2 -\left( 
{R-R_0 } \right)^2} \right)}{2\,R^2}
\end{equation}
In Figure \ref{pot1} the individual components of (\ref{eq22}) are depicted together with 
their superposition and the analytic expression (\ref{eq23}). The agreement between 
the numerical integrals (\ref{eq22}) and the exact expression (\ref{eq23}) is perfect within 
numerical accuracy. It is also observed that the individual contributions in 
(\ref{eq22}) are finite outside the torus (which extends from 
$R=.5\;\mbox{to}\;\mbox{1.5})$, but cancel perfectly when they are 
superimposed. This result confirms the validity of Helmholtz's theorem for 
the chosen case.

\vspace{.3cm}

Next we consider the second term in (\ref{eq15}) which we choose finite on the torus 
surface. Assuming a magnetic field of the form
\begin{equation}
\label{eq24}
B_\vartheta =\frac{C\,r}{R}\;,\quad B_r =0
\end{equation}
we obtain with (\ref{eq4}) the toroidal component of the current density:
\begin{equation}
\label{eq25}
j_\varphi =\frac{C\left( {R_0 +R} \right)}{\mu _0 \,R^2}
\end{equation}
and with (\ref{eq15}) the magnetic force component:
\begin{equation}
\label{eq26}
F_r =-\frac{C^2\,r\,\left( {R_0 +R} \right)}{\mu _0 \,R^3}
\end{equation}
Taking divergence and rotation of this expression we obtain with (\ref{eq11}) and 
(\ref{eq13}) the integral representation of the potentials:
\begin{equation}
\label{eq27}
p\left( {R,\,Z} \right)=\frac{C^2}{4\pi \,\mu _0 }\int\limits_0^{2\pi } 
{\int\limits_0^{2\pi } {\int\limits_0^{r_b } {\frac{1}{R'\,r}\frac{\partial 
}{\partial r}\left( {\frac{R'\,r^2\,\left( {R_0 +R'} \right)}{R'^3}} 
\right)} } } \frac{R'\,r\,dr\,d\vartheta \,d\varphi }{\left| {\vec {x}-\vec {x}'} \right|}
\end{equation}
\begin{equation}
\label{eq28}
u_\varphi \left( {R,\,Z} \right)=\frac{C^2}{4\pi \,\mu _0 
}\int\limits_0^{2\pi } {\int\limits_0^{2\pi } {\int\limits_0^{r_b } 
{\frac{1}{r}\frac{\partial }{\partial \vartheta }\left( {\frac{\,r\,\left( 
{R_0 +R'} \right)}{R'^3}} \right)} } } \frac{\cos \varphi 
\,\,R'\,r\,dr\,d\vartheta \,d\varphi }{ \left| {\vec {x}-\vec {x}'} \right|}
\end{equation}
The radial force in the midplane may be derived from these integrals with 
(\ref{eq22}) and compared with the analytic expression:
\begin{equation}
\label{eq29}
F_R =\frac{C^2\,\left( {R^2-R_0^2 } \right)}{\mu _0 \,R^3}
\end{equation}
\begin{figure}[htbp]
\centerline{\includegraphics[width=5.0in,height=4.0in]{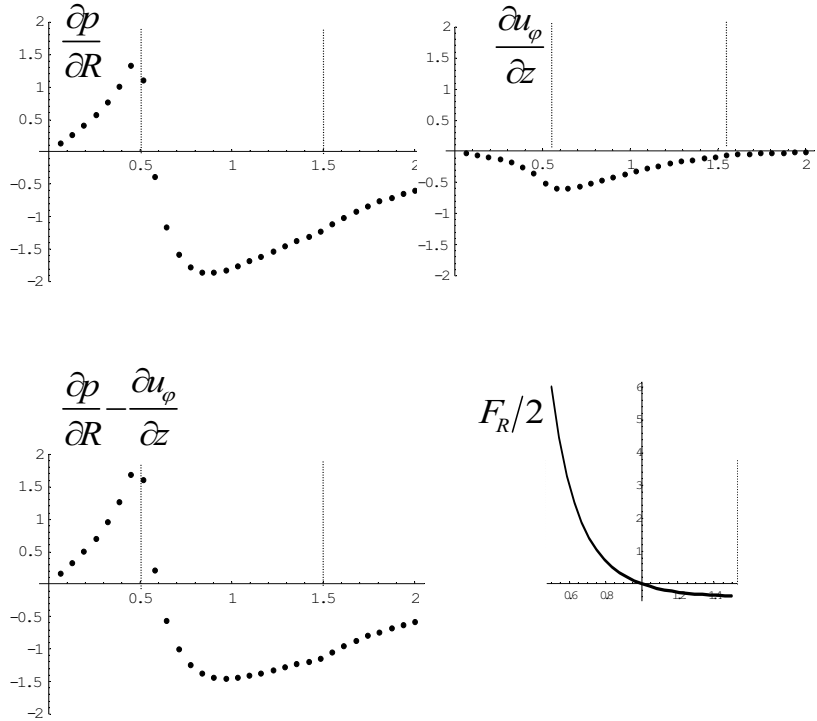}}
\caption{Radial magnetic force $F_R =-j_\varphi \,B_\vartheta $ represented 
by potentials}
\label{pot2}
\end{figure}
In Figure \ref{pot2} we show the result. Obviously, the force as derived from the 
potentials (\ref{eq27}) and (\ref{eq28}) bears little resemblance with the given force $F_R 
$ (note that the scale of the $F_R $ - plot has been reduced by a factor of 2 
in Fig. 3). Furthermore, the superposition of the derivatives of the 
potentials does by no means vanish outside the torus. As already mentioned, 
this behaviour is to be expected for integrals like (\ref{eq27}) and (\ref{eq28}), in 
particular, when the force is finite on the boundary as in (\ref{eq26}). This holds also for the electric force -- the first term in (\ref{eq1}) -- when the body carries a total charge resulting in a finite boundary field. Helmholtz's theorem is apparently not applicable when the vector field is discontinuous 
which is a common property of the electromagnetic forces except in specially 
constructed cases like (\ref{eq16}).

Our investigation so far reveals that the Helmholtz formalism for 
determining the potentials from which the electromagnetic forces could be 
derived will fail in most practical instances. On the other hand, physical 
differential equations like (\ref{eq8}) or (\ref{eq9}) can only have a solution, when the 
electric potential or the pressure do exist. It is therefore of great 
importance to explore different methods which seek a solution for the 
potentials. The next Section will deal with standard methods being used in 
plasma physics for obtaining solutions to the potential problem of the 
electromagnetic forces.
 
\vspace{.5cm}

\textbf{\large 3 Standard methods for solving first order vector field equations 
in plasma physics}

\vspace{.3cm}

\textbf{\large 3.1 Ohm's law in a plasma}

In an early paper of 1958 Kruskal and Kulsrud \cite{3} examined the conditions 
under which a plasma can be confined by a magnetic field. They realized that 
Ohm's law (\ref{eq9}) leads in conjunction with Faraday's law and the assumption of 
a steady state to a special first order equation $-\vec {B}\cdot \nabla \Phi 
=\eta \,\vec {j}\cdot \vec {B}$, since the electric field should be 
expressible as the gradient of a scalar, if $rot\,\vec {E}=0$. For this type 
of equation they coined the term `magnetic differential equation'. Newcomb 
\cite{4} formulated a solvability criterion $\oint {\left( {{\eta \,\vec {j}\cdot 
\vec {B}} \mathord{\left/ {\vphantom {{\eta \,\vec {j}\cdot \vec {B}} B}} 
\right. \kern-\nulldelimiterspace} B} \right)} \,dl=0$, where the 
integration has to be carried out along closed magnetic field lines. Pfirsch 
and Schl\"{u}ter \cite{5} made an attempt to solve the magnetic differential 
equation without taking reference to Helmholtz's theorem, although in effect 
they were trying to determine the gradient part of (\ref{eq9}). They used a magnetic 
model field in the geometry of Figure 1 where the poloidal field lines 
closed themselves on circles. This field satisfied Maxwell's equation (\ref{eq5}), 
but not exactly (\ref{eq4}) and the force equilibrium condition (\ref{eq8}). Nevertheless, 
they obtained an estimate for the plasma diffusion coefficient in a toroidal 
configuration. Later on Maschke \cite{6} generalized their results for a true 
equilibrium configuration that was supposed to satisfy both (\ref{eq4}) and (\ref{eq8}). He 
was able to confirm the order of magnitude of the predicted plasma losses. 
The present state of the art to calculate collisional transport in tokamaks 
including the effect of trapped particles can be found, e.g., in \cite{7}.

Following the method originally adopted in \cite{6} we assume that an 
axisymmetric toroidal field configuration exists (e.g. `tokamak' \cite{8}), 
where the poloidal magnetic field components may be derived from the 
toroidal component of a vector potential:
\begin{equation}
\label{eq30}
B_R =-\frac{1}{R}\frac{\partial \psi }{\partial Z}\;,\quad B_Z 
=\frac{1}{R}\frac{\partial \psi }{\partial R}\;,\quad \psi =R\,A_\varphi 
\end{equation}
The poloidal field lines follow the contours of the surfaces $\psi 
=\mbox{const}$ which are nested, but not necessarily concentric circles as in 
Figure \ref{torus}. The inner product of (\ref{eq9}) with the magnetic field yields the magnetic 
differential equation: 
\begin{equation}
\label{eq31}
B_R \,E_R +B_Z \,E_Z =\eta \,\vec {j}\cdot \vec {B}-E_\varphi \,B_\varphi =s
\end{equation}
In steady state the poloidal electric field components may be expressed as 
the gradient of a potential, whereas the toroidal component $E_\varphi =U 
\mathord{\left/ {\vphantom {U R}} \right. \kern-\nulldelimiterspace} R$ 
could be induced by a transformer that produces a loop voltage $U$. 
The authors of \cite{3, 4, 5, 6} assumed that (\ref{eq31}) has a solution which means that a 
potential exists such that one can write: 
\begin{equation}
\label{eq32}
B_R \,\frac{\partial \phi }{\partial R}+B_Z \,\frac{\partial \phi }{\partial 
Z}=-s
\end{equation}
This equation specifies the derivative of the potential along the poloidal 
field lines. Straightforward integration over the contours of constant flux 
yields the `potential'
\begin{equation}
\label{eq33}
\phi =\phi _0 \left( \psi \right)-\int\limits_{l_0 }^l {\frac{R\,s}{\left| 
{\nabla \psi } \right|}} \,dl
\end{equation}
where $dl$ is a line element on a contour $\psi =\mbox{const}$. Newcomb's 
criterion imposes the integrability condition
\begin{equation}
\label{eq34}
\oint\limits_{\psi =\mbox{const}} {\frac{R\,s}{\left| {\nabla \psi } 
\right|}} \,dl\,=0
\end{equation}
which must be satisfied by $s$ in order to avoid a multi-valued potential. 

Although the described procedure yields a function $\phi \left( {\psi ,\,l} 
\right)$ in a `flux coordinate system', one cannot be certain whether it 
represents a potential which exists independently of the chosen coordinate 
system. This would be guaranteed if we had obtained the potential from the 
solution of a Poisson equation like (\ref{eq11}), but this was not possible, as we 
do not have -- at this point -- any information on the velocity field that 
determines the Laplacian of the potential according to (\ref{eq9}):
\begin{equation}
\label{eq35}
\Delta \phi =\vec {B}\cdot rot\,\vec {v}-\vec {v}\cdot rot\,\vec {B}
\end{equation}
There is only some knowledge on the current and field configuration which 
enters into $s$ as defined by (\ref{eq31}). Maybe this was the reason why the 
Helmholtz method was ignored in the analysis of plasma equilibria.

In order to guarantee the existence of the potential in any coordinate 
system, we must satisfy the necessary and sufficient condition $rot\,\vec 
{E}=0$, or:
\begin{equation}
\label{eq36}
\frac{\partial E_R }{\partial Z}=\frac{\partial E_Z }{\partial R}
\end{equation}
We can apply this condition when we take the gradient of (\ref{eq31})
\begin{eqnarray}
\label{eq37}
 B_R \frac{\partial E_R }{\partial R}+B_Z \frac{\partial E_Z }{\partial 
R}+E_R \frac{\partial B_R }{\partial R}+E_Z \frac{\partial B_Z }{\partial 
R}=\frac{\partial s}{\partial R} \nonumber\\ \\
 B_R \frac{\partial E_R }{\partial Z}+B_Z \frac{\partial E_Z }{\partial 
Z}+E_R \frac{\partial B_R }{\partial Z}+E_Z \frac{\partial B_Z }{\partial 
Z}=\frac{\partial s}{\partial Z} \nonumber
 \end{eqnarray}
and derive with (\ref{eq36}) and (\ref{eq5}) two magnetic differential equations for the 
electric field components: 
\begin{equation}
\label{eq38}
\vec {B}\cdot \nabla \left( {\frac{E_R }{R\,B_Z }} 
\right)=\frac{1}{R}\frac{\partial }{\partial R}\left( {\frac{s}{B_Z }} 
\right)\;,\quad \vec {B}\cdot \nabla \left( {\frac{E_Z }{R\,B_R }} 
\right)=\frac{1}{R}\frac{\partial }{\partial Z}\left( {\frac{s}{B_R }} 
\right)
\end{equation}
The solutions are in analogy to (\ref{eq33}):
\begin{eqnarray}
\label{eq39}
 E_R =R\,B_Z \left[ {f_R \left( \psi \right)+I_R \left( {\psi ,\,l} \right)} 
\right]\;,\quad I_R \left( {\psi ,\,l} \right)=\int\limits_{l_0 }^l 
{\frac{\partial }{\partial R}\left( {\frac{s}{B_Z }} \right)} 
\frac{dl}{\left| {\nabla \psi } \right|} \nonumber\\ 
 E_Z =R\,B_R \left[ {f_Z \left( \psi \right)+I_Z \left( {\psi ,\,l} \right)} 
\right]\;,\quad I_Z \left( {\psi ,\,l} \right)=\int\limits_{l_0 }^l 
{\frac{\partial }{\partial Z}\left( {\frac{s}{B_R }} \right)} 
\frac{dl}{\left| {\nabla \psi } \right|}
 \end{eqnarray}
where Newcomb's condition (\ref{eq34}) must also be satisfied. 

Instead of cylindrical coordinates one could also have used spherical 
coordinates with $R=r\,\sin \theta \,,\;Z=r\,\cos \theta \,,\;B_R =B_r 
\,\sin \theta +B_\theta \,\cos \theta ,\;B_Z =\,B_r \,\cos \theta -B_\theta 
\sin \theta \,$. Equation (\ref{eq31}) transforms then into:
\begin{equation}
\label{eq40}
B_r \,E_r +B_\theta \,E_\theta =s
\end{equation}
and the steady state condition $rot\,\vec {E}=0$ becomes:
\begin{equation}
\label{eq41}
\frac{\partial E_r }{\partial \theta }=\frac{\partial \left( {r\,E_\theta } 
\right)}{\partial r}
\end{equation}
Taking now the gradient of (\ref{eq40}) one obtains with (\ref{eq5}) and (\ref{eq41}) the magnetic 
differential equations for the field components in the form:
\begin{equation}
\label{eq42}
\vec {B}\cdot \nabla \left( {\frac{E_r }{R\,B_\theta }} 
\right)=\frac{1}{R\,r}\frac{\partial }{\partial r}\left( 
{\frac{r\,s}{B_\theta }} \right)\;,\quad \vec {B}\cdot \nabla \left( 
{\frac{E_\theta }{R\,B_r }} \right)=\frac{1}{R\,r}\frac{\partial }{\partial 
\theta }\left( {\frac{s}{B_r }} \right)
\end{equation}
with the solutions:
\begin{eqnarray}
\label{eq43}
 E_r =R\,B_\theta \left[ {f_r \left( \psi \right)+I_r \left( {\psi ,\,l} 
\right)} \right]\;,\quad I_r \left( {\psi ,\,l} \right)=\int\limits_{l_0 }^l 
{\frac{1}{r}\frac{\partial }{\partial r}\left( {\frac{s\,r}{B_\theta }} 
\right)} \frac{dl}{\left| {\nabla \psi } \right|} \nonumber \\ 
 E_\theta =R\,B_r \left[ {f_\theta \left( \psi \right)+I_\theta \left( {\psi 
,\,l} \right)} \right]\;,\quad I_\theta \left( {\psi ,\,l} 
\right)=\int\limits_{l_0 }^l {\frac{1}{r}\frac{\partial }{\partial \theta 
}\left( {\frac{s}{B_r }} \right)} \frac{dl}{\left| {\nabla \psi } \right|} 
\end{eqnarray}
The same expressions must result, if we transform the solutions (\ref{eq39}) 
directly into spherical coordinates:
\begin{eqnarray}
\label{eq44}
 E_r \,\sin \theta +E_\theta \,\cos \theta =R\,\left( {B_r \,\cos \theta 
-B_\theta \sin \theta \,} \right)\left[ {f_R \left( \psi \right)+I_R \left( 
{\psi ,\,l} \right)} \right]\; \nonumber \\ \\
 E_r \,\cos \theta -E_\theta \sin \theta \,=R\,\left( {B_r \,\sin \theta 
+B_\theta \,\cos \theta } \right)\left[ {f_Z \left( \psi \right)+I_Z \left( 
{\psi ,\,l} \right)} \right]\; \nonumber
\end{eqnarray}
or:
\begin{eqnarray}
\label{eq45}
\!\!\!E_r \!=\!R\, B_r \sin \theta \cos \theta \left[ {f_R +I_R +f_Z +I_Z } 
\right]+R\, B_\theta \!\left[ {\cos ^2\theta \left( {f_Z +I_Z } \right)-\sin 
^2\theta \left( {f_R +I_R } \right)} \right] \nonumber \\ \\
\!\!\!E_\theta \!=\!R\, B_r \!\left[ {\cos ^2\theta \left( {f_R +I_R } \right)-\sin 
^2\theta \left( {f_Z +I_Z } \right)} \right]-R\, B_\theta \sin \theta \cos 
\theta \left[ {f_R +I_R +f_Z +I_Z } \right] \nonumber 
\end{eqnarray}
In order to make (\ref{eq45}) compatible with (\ref{eq43}) it is apparently necessary to 
require:
\begin{equation}
\label{eq46}
f_R +I_R +f_Z +I_Z =0
\end{equation}
On the other hand, if one substitutes the solution (\ref{eq39}) into (\ref{eq31}), one 
obtains:
\begin{equation}
\label{eq47}
R\,B_R \,B_Z \left( {f_R +I_R +f_Z +I_Z } \right)=s
\end{equation}
The inhomogeneous part $s$ of the magnetic differential equation (\ref{eq32}) must 
apparently vanish, if the function $\phi $ is to exist as a potential 
independent of the coordinate system. This means, of course, that a scalar 
potential for the arbitrarily given vector field $\eta \,\vec {j}-\vec 
{v}\times \vec {B}$ in (\ref{eq9}) does not exist. Only a function $\phi =\phi _0 
\left( \psi \right)$ would satisfy Ohm's law (\ref{eq32}) when $s=0$, but one could 
not consider it as a true potential either. This will be discussed in the 
following subsection.

\vspace{.5cm}

\textbf{\large 3.2 The force equilibrium in a plasma}

The stationary force balance in a magnetized plasma cannot be easily 
satisfied. In fact, there are not any analytic solutions of (\ref{eq8}) known, 
except in the axisymmetric case \cite{9, 10}. From the toroidal component of (\ref{eq8}), namely 
$j_Z B_R -j_R B_Z =0$, follows with (\ref{eq4}) a homogeneous magnetic differential 
equation for the toroidal field component:
\begin{equation}
\label{eq48}
\vec {B}\cdot \nabla \left( {R\,B_\varphi } \right)=0
\end{equation}
and the inner product of (\ref{eq8}) with the magnetic field leads to another 
homogeneous magnetic differential equation 
\begin{equation}
\label{eq49}
\vec {B}\cdot \nabla p=0
\end{equation}
Inserting the obvious solutions: 
\begin{equation}
\label{eq50}
B_\varphi ={F\left( \psi \right)} \mathord{\left/ {\vphantom {{F\left( \psi 
\right)} {R\;,\quad }}} \right. \kern-\nulldelimiterspace} {R\;,\quad 
}p=p\left( \psi \right)
\end{equation}
into the poloidal components of (\ref{eq8}) one finds with (\ref{eq4}) an 
expression for the toroidal component of the current density:
\begin{equation}
\label{eq51}
j_\varphi =R\frac{dp}{d\psi }+\frac{F}{\mu _0 \,R}\frac{dF}{d\psi }
\end{equation}
Substitution into the toroidal component of (\ref{eq4}) yields with (\ref{eq30}) the famous 
nonlinear L\"{u}st-Schl\"{u}ter-Grad-Rubin-Shafranov equation \cite{11}:
\begin{equation}
\label{eq52}
\frac{\partial ^2\psi }{\partial R^2}-\frac{1}{R}\frac{\partial \psi 
}{\partial R}+\frac{\partial ^2\psi }{\partial Z^2}\equiv \Delta ^\ast \psi 
=-\mu _0 \,R^2\frac{dp}{d\psi }-F\,\frac{dF}{d\psi }
\end{equation}
It is the basis for calculating numerically axisymmetric plasma equilibria. 
Recently the equation has been modified \cite{12} to include approximately the 
effect of magnetic islands which break axisymmetry.

One must expect, however, that a function $p\left( \psi \right)$ which 
depends on a single vector component, namely $R\,\vec {e}_R \times A_\varphi 
\,\vec {e}_\varphi $, cannot represent a physical pressure that must be 
defined independent of the coordinate system. In order to see this we 
consider an analytical `Soloviev solution' \cite{9} 
\begin{equation}
\label{eq53}
\psi =\frac{1}{2}\left( {c_0 R^2+b\,R_0^2 } \right)Z^2+\frac{a-c_0 
}{8}\left( {R^2-R_0^2 } \right)^2\;,\quad F=\sqrt {F_0^2 -2\,b\,R_0^2 \, \psi } 
\end{equation}
which satisfies (\ref{eq52}) by construction. The pressure is given as a linear 
function of $\psi $: 
\begin{equation}
\label{eq54}
p\left( {R,\,Z} \right)=p_0 -\left( {a \mathord{\left/ {\vphantom {a {\mu _0 
}}} \right. \kern-\nulldelimiterspace} {\mu _0 }} \right)\,\psi =p_0 -\left( 
{a \mathord{\left/ {\vphantom {a {\mu _0 }}} \right. 
\kern-\nulldelimiterspace} {\mu _0 }} \right)\,R\,A_\varphi \left( {R,\,Z} 
\right)
\end{equation}
Transforming this equation into a Cartesian coordinate system yields:
\begin{equation}
\label{eq55}
p\left( {x,\,y,\,z} \right)=p_0 -\left( {a \mathord{\left/ {\vphantom {a 
{\mu _0 }}} \right. \kern-\nulldelimiterspace} {\mu _0 }} \right)\,\left( 
{x\,A_y -y\,A_x } \right)
\end{equation}
Rotation of the coordinate system around the $y$ - axis by an angle $\alpha$ according to the 
transformation rules
\begin{eqnarray}
\label{eq56}
 x=x'\cos \alpha -z'\sin \alpha \;,\quad z=x'\sin \alpha +z'\cos 
\alpha \;,\quad y=y' \quad \quad \nonumber \\ \\
 A_x =A_{x'} \cos \alpha -A_{z'} \sin \alpha \;,\quad A_z =A_{x'} \sin 
\alpha +A_{z'} \cos \alpha \;,\quad A_y =A_{y'} \nonumber
\end{eqnarray}
results in a pressure field
\begin{equation}
\label{eq57}
p\left( {x',\,y',\,z'} \right)=p_0 -\left( {a \mathord{\left/ {\vphantom {a 
{\mu _0 }}} \right. \kern-\nulldelimiterspace} {\mu _0 }} \right)\,\left[ 
{\left( {x'\,A_{y'} -y'\,A_{z'} } \right)\cos \alpha +\left( {y'\,A_{z'} 
-z'\,A_{y'} } \right)\sin \alpha } \right]\,
\end{equation}
which depends not only on the new coordinates, but also on the rotational 
angle $\alpha $. It is, therefore, not a scalar in the usual definition: 
$p'\left( {\vec {x}'} \right)=p\left( {\vec {x}} \right)\;\mbox{for}\;\vec 
{x}'=\vec {x}$. Our conclusion is then that the magnetic force $\vec 
{j}\times \vec {B}$ does not have a scalar potential which could be 
identified with a physical pressure.

The so-called `force-free' configuration
\begin{equation}
\label{eq58}
\vec {j}\times \vec {B}=\vec {0}
\end{equation}
which plays a role in very diluted astrophysical plasmas, deserves special 
mentioning. Obviously, it makes little sense to replace the $\vec {0}$ - 
vector in (\ref{eq58}) by a gradient and a curl. Whether a solution to (\ref{eq58}) exists, 
must be examined in a different way. In the axisymmetric case we have from 
(\ref{eq52}):
\begin{equation}
\label{eq59}
\Delta ^\ast \psi =-F\,\frac{dF}{d\psi }\equiv -g\left( \psi \right)
\end{equation}
Applying Stokes's theorem on this equation by integrating the toroidal 
current density over the area enclosed by a magnetic surface one has:
\begin{equation}
\label{eq60}
\oint {\frac{\left| {\nabla \psi } \right|}{R}} 
\,dl=\mathop{{\int\!\!\!\!\!\int}\mkern-21mu \bigcirc} {g\left( \psi 
\right)} \frac{dR\,dZ}{R}
\end{equation}
With the inverse function
\begin{equation}
\label{eq61}
\psi =\int {\frac{F}{g\left( F \right)}} \,dF\;,\quad \nabla \psi 
=\frac{F}{g}\,\nabla F
\end{equation}
one obtains from (\ref{eq59}):
\begin{equation}
\label{eq62}
\Delta ^\ast F=-\frac{g^2}{F}-\frac{g}{F}\left| {\nabla F} 
\right|^2\frac{d}{dF}\left( {\frac{F}{g}} \right)
\end{equation}
Stokes's theorem applied on this equation gives:
\begin{equation}
\label{eq63}
\oint {\frac{\left| {\nabla F} \right|}{R}} 
\,dl=\mathop{{\int\!\!\!\!\!\int}\mkern-21mu \bigcirc} {\left( 
{\frac{g^2}{F}+\frac{g}{F}\left| {\nabla F} \right|^2\frac{d}{dF}\left( 
{\frac{F}{g}} \right)} \right)} \frac{dR\,dZ}{R}
\end{equation}
Substitution of (\ref{eq61}) into (\ref{eq60}) yields on the other hand:
\begin{equation}
\label{eq64}
\frac{F}{g}\oint {\frac{\left| {\nabla F} \right|}{R}} 
\,dl=\mathop{{\int\!\!\!\!\!\int}\mkern-21mu \bigcirc} g \frac{dR\,dZ}{R}
\end{equation}
Elimination of the line integral over the poloidal current density on the 
left-hand-sides of (\ref{eq63}) and (\ref{eq64}), and using (\ref{eq61}) again results in an 
integral equation
\begin{equation}
\label{eq65}
g\,\mathop{{\int\!\!\!\!\!\int}\mkern-21mu \bigcirc} g 
\frac{dR\,dZ}{R}=F\mathop{{\int\!\!\!\!\!\int}\mkern-21mu \bigcirc} {\left( 
{\frac{g^2}{F}+\frac{1}{F^3}\left( {g^2-F\frac{dg}{d\psi }} \right)\left| 
{\nabla \psi } \right|^2} \right)} \frac{dR\,dZ}{R}
\end{equation}
which can only be satisfied for $g=F{dF} \mathord{\left/ {\vphantom {{dF} 
{d\psi =0}}} \right. \kern-\nulldelimiterspace} {d\psi =0}$. This may be 
demonstrated by choosing $g=\mbox{const}$ and performing a partial 
integration on (\ref{eq65}):
\begin{equation}
\label{eq66}
g^2\!\!\!\!\!\oint\limits_{\psi =\mbox{const}}\!\!\!\!\! {\frac{Z\left( R,\,\psi \right)dR\,}{R}} 
\,=F\,\frac{g^2}{F}\!\!\!\!\!\oint\limits_{\psi =\mbox{const}}\!\!\!\!\! {\frac{Z\left( R,\,\psi 
\right)dR\,}{\,R}} +F\mathop{{\int\!\!\!\!\!\int}\mkern-21mu \bigcirc} 
{\left( {\frac{g^3\,Z}{F^3}\frac{\partial \psi }{\partial 
Z}+\frac{g^2}{F^3}\left| {\nabla \psi } \right|^2} \right)} 
\frac{dR\,dZ}{R}
\end{equation}
The first term on the right-hand-side cancels against the left-hand-side, 
whereas the second term cannot vanish on every surface except for $g=0$. 
This may be verified by adopting the force-free Soloviev solution (\ref{eq53}) with 
$a=0$ and $g=-b\,R_0^2 $. We must conclude then that equation (\ref{eq58}) in 
conjunction with Ampere's law (\ref{eq4}) does not have a solution.

\vspace{.5cm}

\textbf{\large 4 Discussion and conclusion }

Our attempts to determine scalar and vector potentials for the 
electromagnetic forces were only partially successful. Thanks to Helmholtz's 
theorem the task of formulating and calculating the potentials should be an 
easy exercise, but it was found that in most practical cases -- where the 
forces are discontinuous across the boundary of the conductors -- the 
theorem is not applicable.

On the other hand, the potentials must exist, if the electromagnetic forces 
are to be balanced against mechanical forces in a stationary state, in 
particular in fluid conductors like plasmas, for example. We have examined 
the usual method for finding the electrostatic potential in Ohm's law and 
the pressure potential in the force balance of a plasma without having 
regard to Helmholtz's theorem. Although it appears possible to find 
solutions in specially chosen coordinate systems, the functions from which 
the forces may be derived cannot be identified with physical potentials, as 
they are not invariant against arbitrary transformations of the coordinate 
system.

Our finding has important implications on the physics of magnetic plasma 
confinement. If it is not possible to find solutions for the steady state 
equations, a plasma can only be confined in a `quasistationary' state at 
best. This means that the partial time derivatives neither in the law of 
induction, nor in the equations of motion can be neglected. The plasma may 
develop into a fluctuating turbulent state where strict axisymmetry does not 
prevail anymore. The observed `anomaly' of plasma losses is probably due to 
this lack of a well balanced equilibrium in confining devices like tokamaks, 
stellarators, reversed field pinches etc. Harold Grad -- one of the authors 
of equation (\ref{eq52}) -- came in his last paper \cite{13} to a similar conclusion: 
``The discovered lack of pressure balance is not a special consequence of 
one idealized model but carries over broadly to a variety of physical 
refinements and generalizations. Indeed, the mathematical result has 
profound physical consequences which have gradually entered the field in the 
context of rational rotation number resonances and multiple helicities, 
island formation, and turbulence.''

%\end{large}
\end{document}